\documentclass[a4paper]{article}

\usepackage[dvips]{feynmp}

{\def\$#1: #2 ${#2}}

\usepackage{amsmath,shortvrb}

\newcommand{\meta}[1]{$\langle$\textit{#1}$\rangle$}
\newcommand{\connected}{\mathrel{\ast}}
\newcommand{\constraint}{\mathrel{\sim}}
\newenvironment{commands}%
 {\begin{list}{}%
   {\setlength{\leftmargin}{2em}%
    \setlength{\rightmargin}{2em}%
    \setlength{\itemindent}{-1em}%
    \setlength{\listparindent}{0pt}%
    }}%
 {\end{list}}

\providecommand{\preprintno}[1]{}

\IfFileExists{mflogo.sty}%
  {\usepackage{mflogo}%
   \newcommand{\FMF}{\texttt{feyn}\textlogo{MF}}}%
  {\newcommand{\MF}{\textsf{META}\-\textsf{FONT}}%
   \newcommand{\MP}{\textsf{META}\-\textsf{POST}}%
   \newcommand{\FMF}{\texttt{feyn}\textsf{MF}}}

\AtBeginDocument{%
  \DeclareFontFamily{OT1}{ccr}{}
  \DeclareFontShape{OT1}{ccr}{m}{it}{%
    <7><10><10.95><12>ccti10}{}}

\begin{document}
\bibliographystyle{prsty}
\MakeShortVerb{\|}

\title{Drawing Feynman Diagrams with \LaTeX{} and \MF}

\author{%
  Thorsten Ohl\thanks{e-mail:
    \texttt{Thorsten.Ohl@Physik.TH-Darmstadt.de}}\\
  \hfil \\
  Technische Hochschule Darmstadt \\
  Schlo\ss gartenstr. 9 \\
  D-64289 Darmstadt \\
  Germany}

\preprintno{IKDA 95/20}
\date{May 1995}

\newcommand{\Version}{1.0}
\newcommand{\Date}{May 1995}


\maketitle
\begin{abstract}
  \FMF{} is a \LaTeX{} package for easy drawing of professional
  quality Feynman diagrams with \MF{} (or \MP).  \FMF{} lays out most
  diagrams satisfactorily from the structure of the graph without
  any need for manual intervention.  Nevertheless all the power of
  \MF{} (or \MP) is available for the most complicated cases.
\end{abstract}

\newpage
\begin{fmffile}{diagrams}
\section{Introduction}
\label{sec:introduction}

In recent years, \TeX~\cite{Knu86a} and \LaTeX~\cite{Lam94} (or other
macro packages for structured markup on top of \TeX) have
revolutionized the way we share information in theoretical physics
(and other areas).  Not only does \TeX{} provide typographical
capabilities which transcend those of commercial ``word processors''
substantially, \TeX{} documents are also completely portable among
computer systems.  Since
implementations are available for essentially all computers in use in
the physics community,
documents can be shared without the usual restrictions
of proprietary data formats.  This has enabled us to collaborate on
papers with colleagues on the other side of the globe, to replace the
mailing of hard copy preprints by electronic transmission and to
submit these papers electronically to the publisher.

\TeX's portability comes with a price, though.  It does deliberately
not address the issue of graphical information, apart from the
rudimentary (but very useful) capabilities of the \LaTeX{}
|picture| environment and similar packages~\cite{GMS94}.
More complex graphics can only be handled by inclusion of more or less
device dependent external graphics files.

More recently, the inclusion of graphics files in the
PostScript~\cite{Ado90} page description language has emerged as a de
facto standard.  This approach restricts the portability of documents
to installations were PostScript printers (or emulators) are
available.  The popularity of such devices makes this an almost
moot point, though.

Nevertheless, handling graphics in an environment completely different
from the (\TeX) text environment causes other problems.  Some popular
packages that employ graphical user interfaces will force PostScript
fonts for labeling on the user.  These fonts will usually not blend
smoothly with other fonts used in text and equations.  More
importantly, these packages usually lack the ability to create complex
mathematical expressions which would be useful in the labels of
figures in physics papers.  Finally,
these tools are usually less than portable to the extent that changing
jobs means changing tools.

Currently there are a couple of tools available that address one or
more of the above points in the context of drawing Feynman diagrams,
which form one of the most frequent classes of graphics in physics
papers.  Michael Levine's |feynman| package~\cite{Lev90} is
implemented on top of the standard \LaTeX{} |picture|
environment~\cite{Lam94}.  This makes it completely portable, but the
graphics output is less than perfect.  This is not the fault of the
|feynman| package, but rooted in principle limitations of the
|picture| environment.  Jos Vermaseren's |axodraw|
package~\cite{Ver94} uses \verb+\special+ to access PostScript
primitives for drawing diagrams.  This approach is inherently not
portable (the mentioned ubiquity of PostScript printers makes this a
minor point, though) but very flexible and produces substantially more
pleasant graphics.  Both packages take no advantage of the formal
structure of Feynman graphs, but require the user to specify the
layout manually using low level graphics primitives.

It is possible to go one step further and move from low-level
tools working on points and curves to a high-level markup system
working on the mathematical structure of graphs.  This step will free
the user from having to think about the layout and allow him to
concentrate on the structure of the graph instead.

In this paper I will describe such a system, \FMF, which is
\emph{completely} portable among \TeX{} installations.  It is unique
among packages for drawing Feynman diagrams in \emph{combining} the
following features:
\begin{itemize}
  \item Simplicity and conciseness for common diagrams.  The
    scattering diagram in figure~\ref{fig:simple} can be specified
    \emph{completely} in five lines of \LaTeX:
    \begin{verbatim}
      \begin{fmfchar*}(40,30) \fmfpen{thick}
        \fmfleft{i1,i2} \fmfright{o1,o2}
        \fmf{fermion}{i1,v1,o1} \fmf{fermion}{i2,v2,o2}
        \fmf{photon,label=$q$}{v1,v2} \fmfdot{v1,v2}
      \end{fmfchar*}\end{verbatim} 
    \begin{figure}[t]
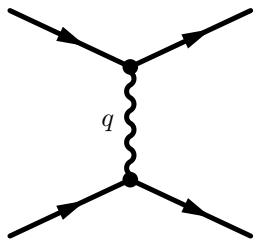

      \begin{center}
      \begin{fmfchar*}(40,30) \fmfpen{thick}
        \fmfleft{i1,i2} \fmfright{o1,o2}
        \fmf{fermion}{i1,v1,o1} \fmf{fermion}{i2,v2,o2}
        \fmf{photon,label=$q$}{v1,v2} \fmfdot{v1,v2}
      \end{fmfchar*}
      \end{center}
      \caption{\label{fig:simple}Simple scattering diagram.}
    \end{figure}
    It is never necessary to draft the diagram on graph paper or to
    perform calculations to determine the position of vertices
    manually.
  \item Expressiveness for arbitrarily complex diagrams (see the
    examples below).
  \item Extensibility.
  \item Portability.  No graphics devices are needed beyond a standard
    \TeX{} installation.
  \item Arbitrary \TeX-labels.
\end{itemize}

The paper is structured as follows: I begin by describing the design
of \FMF{} in section~\ref{sec:design}. Then I describe some
details of the implementation in
section~\ref{sec:implementation}.  After a brief discussion of the
most important user commands in section~\ref{sec:usage}, I conclude.

\section{Design}
\label{sec:design}

A clear cut distinction between ``design'' and ``implementation'' is
certainly fictitious.  As in most programs with more than several
hundred lines of code, designs have been adapted as implementation
progressed and feedback from early users came in.

\subsection{Goals}

As mentioned in the introduction, \FMF{} was to meet the following
competing design goals:
\begin{itemize}
  \item convenience and ease-of-use,
  \item expressiveness,
  \item extensibility, and
  \item portability.
\end{itemize}
Of these, extensibility is not a goal in itself but should rather be
viewed as a derived
goal. It appears impossible to reconcile ease-of-use with
expressiveness in the straight jacket of an inextensible
implementation.  It is much more effective to provide an extensible
environment where simple building blocks can be used for the
straightforward solution of simple problems \emph{and} can be combined
to solve the most complicated problems, once the software system has
been mastered by the user.

The reconciliation of convenience and expressiveness was made
possible by providing two different modes:
\begin{itemize}
  \item \emph{graph mode}, in which the layout is determined
    automatically from a simple mathematical description of the graph,
    and
  \item \emph{immediate mode}, in which the user has complete freedom
    but at a basic familiarity with \MF{} is recommended.
\end{itemize}

The goal of portability was easily reached by basing the
implementation of \TeX{} and \MF, because both programs will be
available to all potential users of the software.

\subsection{Languages}

The primary user interface is a set of \LaTeX{} macros.  It is
therefore possible to keep the whole paper, including graphics, in a
single file.  This is, among other things, very convenient for
exchanging manuscripts by electronic mail~\cite{Ver94}.  Also, no new
syntax has to be learned by the user.

\MF~\cite{Knu86b} (or alternatively \MP~\cite{Hob92}) has been chosen
as the low level graphics engine for the following reasons:
\begin{itemize}
  \item{} \MF{} is part of any reasonable \TeX{} installation, therefore
    available to all potential users,
  \item{} \MF{} output is readily included in \TeX{} documents in the
    form of (unusual) characters,
  \item{} \MF{} has very powerful graphics primitives~\cite{Knu86b},
    which allow high quality output, and
  \item{} \MF{} has builtin linear algebra~\cite{Knu86b}, which can be
    employed for automatic layout algorithms, as detailed below.
\end{itemize}
Without taking advantage of these features, the implementation in
other languages would have been much more complex.

\subsection{Algorithmic layout}
\label{sec:algorithmic}

Early in the design it was clear that \FMF{} should in at least one
mode of operation accept a \emph{mathematical} description of a graph
and create the layout of the corresponding Feynman diagram
automatically.  It should also not rely on a database of common
topologies, because such a database will necessarily remain incomplete.

Every graph can be specified completely by giving a
set~$A$ of pairs.  The set~$V$ of vertices is then given by
\begin{equation}
  V = \left\{\, v \mid \exists a\in A:
                a = (v,v') \vee a = (v',v) \,\right\} \,.
\end{equation}
The set~$A$ will henceforth be called the set of arcs.  It is
useful to introduce the sets of vertices connected to~$v$
\begin{equation}
  \alpha(v) = \left\{\, v' \in V \mid v \connected v' \,\right\} \,,
\end{equation}
where~$\connected$ denotes the symmetrical relation
\begin{equation}
 \left(v \connected v'\right)
   \Leftrightarrow
     \left(\exists a\in A: a = (v,v') \vee a = (v',v)\right)
\end{equation}
of being connected.

The obvious first candidate for a function that should be minimized is
the sum of the squared lengths of arcs
\begin{equation}
\label{eq:length}
  l(v_1,\ldots,v_n)
    = \sum_{\substack{i,j = 1\\ v_i \connected v_j}}^n (v_i - v_j)^2.
\end{equation}
As is stands,~(\ref{eq:length}) is not yet sufficient,
because~$v_i=\hat v$ for all~$i$ is obviously a
solution corresponding to the minimum~$l=0$ for all~$\hat v$.
In order to lift the
degeneracy, we have to break translational invariance.  However,
breaking the translational invariance by demanding for instance that
the center of gravity~$\sum_{i=1}^n v_i/n$ coincides with the center
of the picture is still not enough, because we hardly want all arcs to
shrink to a point.

It will now be useful to introduce the set of all external vertices
\begin{equation}
\label{eq:Vext}
  V^{\text{ext}} = \left\{\, v\in V \mid \vert\alpha(v)\vert = 1\,\right\}
\end{equation}
and its complement, the set of all internal vertices
\begin{equation}
\label{eq:Vint}
  V^{\text{int}} = V \setminus  V^{\text{ext}}\,.
\end{equation}
{}From~(\ref{eq:length}) we see that the vertices in~$V^{\text{ext}}$
will occupy the same position as their single neighbor, unless their
position is fixed explicitly.  It is therefore necessary to specify
explicit positions for the external vertices.  In the implementation
of \FMF, commands are provided to place a list of external vertices on
``galleries'' along the sides of the diagram.  Using a similar
strategy for external vertices, (\ref{eq:length}) has been used for
example in~\cite{SL94} with good results for automated drawing of tree
diagrams in PostScript.

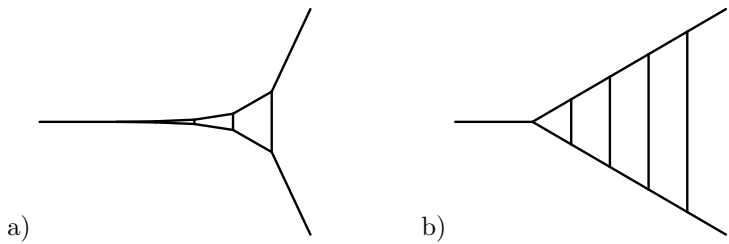
\begin{figure}
  \begin{center}
    a)
    \begin{fmfgraph}(40,30)
      \fmfleft{i} \fmfright{o1,o2}
      \fmf{plain}{i,v}
      \fmf{plain}{o1,d1,d2,d3,d4,v,u4,u3,u2,u1,o2}
      \begin{fmffor}{n}{1}{1}{4}
        \fmf{plain}{d[n],u[n]}
      \end{fmffor}
    \end{fmfgraph}
    \qquad
    b)
    \begin{fmfgraph}(40,30)
      \fmfleft{i} \fmfright{o1,o2}
      \fmf{plain}{i,v}
      \fmf{plain}{o1,d1,d2,d3,d4,v,u4,u3,u2,u1,o2}
      \fmffreeze
      \begin{fmffor}{n}{1}{1}{4}
        \fmf{plain}{d[n],u[n]}
      \end{fmffor}
    \end{fmfgraph}
    \caption{\label{fig:ladders}%
     Ladder diagram, a) using the ``action'' (\ref{eq:length})
     and b) using the improved ``action'' (\ref{eq:wlength}).}
  \end{center}
\end{figure}

While~(\ref{eq:length}) gives fair results for almost all tree
diagrams, it can fail miserably on loop diagrams, as witnessed in
figure~\ref{fig:ladders}a.  A simple generalization of (\ref{eq:length})
can improve results immensely
\begin{equation}
\label{eq:wlength}
  L(v_1,\ldots,v_n)
    = \frac{1}{2} \sum_{\substack{i,j = 1\\ v_i \connected v_j}}^n
            t_{ij} (v_i - v_j)^2\,.
\end{equation}
The elements of the symmetrical ``tension'' matrix~$t_{ij}$ are positive
numbers that default to~$1$ and can be used to tune the layout.

\begin{figure}
  \begin{center}
    \begin{fmfgraph*}(30,30)
      \fmfleft{e1,e2} \fmfright{e4,e3}
      \fmf{plain}{v4,v1,v2,v3}
      \fmf{plain,label=$t_{ij}=4$,label.side=left,tension=4}{v3,v4}
      \begin{fmffor}{n}{1}{1}{4}
        \fmf{plain}{v[n],e[n]}
      \end{fmffor}
    \end{fmfgraph*}
    \qquad
    \begin{fmfgraph*}(30,30)
      \fmfleft{e1,e2} \fmfright{e4,e3}
      \fmf{plain}{v4,v1,v2,v3}
      \fmf{plain,label=$t_{ij}=1$,label.side=left}{v3,v4}
      \begin{fmffor}{n}{1}{1}{4}
        \fmf{plain}{v[n],e[n]}
      \end{fmffor}
    \end{fmfgraph*}
    \qquad
    \begin{fmfgraph*}(30,30)
      \fmfleft{e1,e2} \fmfright{e4,e3}
      \fmf{plain}{v4,v1,v2,v3}
      \fmf{plain,label=$t_{ij}=1/4$,label.side=left,tension=1/4}{v3,v4}
      \begin{fmffor}{n}{1}{1}{4}
        \fmf{plain}{v[n],e[n]}
      \end{fmffor}
    \end{fmfgraph*}
    \caption{\label{fig:tension}%
     Varying the tension parameter.}
  \end{center}
\end{figure}
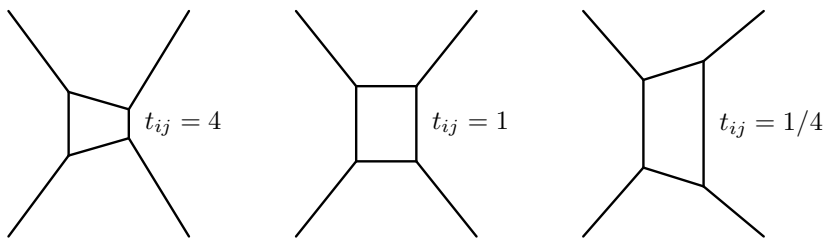

The effect of the tension parameter can be understood by imagining the
graph as consisting of rubber bands.  Changing the tension of an arc
will pull adjacent vertices together or allow them to move apart.  As
an example, figure~\ref{fig:tension} shows the effect of varying the
tension of one line from~$4$ to~$1/4$.

The improved ladder diagram in figure~\ref{fig:ladders}b has been
drawn with vanishing tension of the arcs, which will result in
straight lines for the stems.

In fact, the effect of vanishing tension can also be achieved by
laying out subgraphs step by step.  By freezing the layout of the
subgraph excluding the rungs in figure~\ref{fig:ladders}b first and
adding the rungs later, we arrive at the same result.  Obviously this
procedure can be iterated for graphs of arbitrary complexity.

While there are also commands to fix the position of a vertex or to
shift its position, it turns out that the most effective way of
drawing Feynman diagrams consists in a combination of the stepwise
construction of subgraphs and adjustment of tensions.  User's
discretion is advised in tuning tension parameters.  More often than
not, the defaults give satisfactory results that can be made perfect
by adjusting the tension of a single arc or loop.  Tuning too many
tensions is not likely to improve the results and is almost as time
consuming as choosing the layout manually.

Technically, the most convenient aspect of~(\ref{eq:wlength}) is that
minimizing it leads to \emph{linear} equations (see~(\ref{eq:layout})
below), which are easily solved.  It would in principle be possible to
investigate improved functions like
\begin{equation}
\label{eq:quartic}
  N(v_1,\ldots,v_n)
    = \sum_{\substack{i,j = 1\\ v_i \connected v_j}}^n
            \left( (v_i - v_j)^2 - \delta^2 \right)^2\,,
\end{equation}
which would avoid the problem of figure~\ref{fig:ladders}a to some
extent by favoring arcs of length~$\delta$.  However, the prize in
having to solve a system of non-linear equations is certainly too
high, in particular because it will be impossible to prove that
reasonable results will result from \emph{all} user input.

\subsection{Constraints}

The method of Lagrange multipliers allows us to specify linear
constraints among vertices
\begin{equation}
\label{eq:constraints}
    v_i - v_j = d_{ij}\,,
\end{equation}
while still dealing with linear equations.  Therefore we add the term
\begin{equation}
  \Lambda = \sum_{\substack{i,j = 1\\ v_i \constraint v_j}}^n
        \lambda_{ij}\cdot(v_i - v_j - d_{ij})
\end{equation}
to the ``action''~(\ref{eq:wlength}).  Then the system
of~$2n_\alpha+2n_\gamma$ linear equations determining the layout is
\begin{equation}
\label{eq:layout}
  \begin{split}
    0 & = \frac{\partial}{\partial v_i^\mu} (L+\Lambda)
            = \sum_{\substack{j = 1\\ v_i \connected v_j}}^n
                  t_{ij} (v_i^\mu - v_j^\mu)
               + \sum_{\substack{j = i+1 \\ v_i \constraint v_j}}^n
                    \lambda_{ij}^\mu
               - \sum_{\substack{j = 1 \\ v_i \constraint v_j}}^{i-1}
                    \lambda_{ji}^\mu \\
    0 & = v_i^\mu - v_j^\mu - d_{ij}^\mu\;\; \mid  v_i \constraint v_j\,.
  \end{split}
\end{equation}

Experience shows that non-linear constraints like fixing a
distance
\begin{equation}
    \vert v_i - v_j \vert = \delta
\end{equation}
would be useful sometimes, but as discussed at the end of
section~\ref{sec:algorithmic}, their implementation is beyond the
scope of \FMF.

\subsection{Immediate mode}

In addition to the graph mode for algorithmic layout that has been
described in the previous section, \FMF{} also has an immediate mode
to provide the user with maximum flexibility.  Immediate mode is
particularly useful for unusually curved arcs, which can not be
specified easily in graph mode.  In this mode, arbitrary \MF{} paths
can be drawn, either specified by absolute coordinates or derived from
arcs entered previously in graph mode.  For a detailed description of
\MF's features, the reader is referred to~\cite{Knu86b}.

\subsection{Extension mechanism}
\label{sec:extensions}

A great variety of different line styles is in use in the physics
community.  \FMF{} provides the most common of them per default, as
displayed in table~\ref{tab:line-styles}.  While it would at best be
inefficient to support an exhaustive list of such styles, it would
probably be a futile effort anyway.  Instead, \FMF{} implements an
extension mechanism that allows the user to install custom line styles
of arbitrary complexity.  For this purpose, a macro |style_def| is
provided.  This macro defines a \MF{} function that does the actual
drawing based on the path it receives as an argument.  Furthermore it
records the name of the function to make it available to graph mode
and immediate mode as well.

\section{Implementation}
\label{sec:implementation}

\FMF{} is implemented in the form of two macro packages: |feynmf.sty|
for the \LaTeX{} part and |feynmf.mf| for the \MF{} part.  Let us
consider both of them in turn.

\subsection{\LaTeX{} macros}

Most macros in |feynmf.sty| are trivial in that they are just writing
their \MF{} equivalent to the \MF{} input file.  The |\fmf| macro, for
example, is just the \TeX{} version of the |vconnect| \MF{} function:
\begin{quote}
  |\def\fmf#1#2{\fmfcmd{vconnect ("#1", \pfx{#2});}}|
\end{quote}
Here |\fmfcmd| writes its expanded argument to the \MF{} file and
|\pfx| adds a ``|__|'' prefix to each member of the comma separated
list it takes as an argument.  This measure protects the unwary user
from the mysterious errors caused by accidentally using a \MF{}
reserved word for a vertex name.

A non-trivial aspect of the |\pfx| macro worth mentioning is that it
works by macro expansion alone (in \TeX's ``mouth'' in Knuth's
terminology~\cite{Knu86a})
and does not need to redefine any macro (an operation that would have
to happen in \TeX's ``stomach'' in Knuth's terminology).  This is
necessary for making |\pfx| work inside of a |\write|, where macros
are expanded but redefinitions are prohibited (see~\cite{Knu86a},
Appendix~D).  Traditional implementations of looping constructs
(e.g.~|\loop| \ldots |\repeat|) work by redefining a continuation and
are therefore unavailable inside of a |\write|.  A possible solution
would be
to use a temporary variable and force expansion of |\pfx| outside
of the |\write|.  A far more elegant solution uses a subset of a
partial implementation~\cite{Jef90} of
$\lambda$-calculus~\cite{Chu41} in \TeX's ``mouth''.

The other unusual aspect of the \LaTeX{} macros is the
\begin{quote}
  |\grepfile{|\meta{pattern}|}{|\meta{infile}|}{|\meta{outfile}|}|
\end{quote}
macro that copies all lines starting with |:|\meta{pattern}|:| from
\meta{infile} to \meta{outfile} after stripping off the
|:|\meta{pattern}|:|.  It is used to extract the label information
that the \MF{} macros have stored in the |log|-file.  This trick
overcomes \MF's limitation of not being able to open any other files
than the terminal, the |gf|-file, the |tfm|-file, and  the |log|-file.
The implementation of this macro is straightforward using \TeX's
pattern matching macro definitions.

However, all subtleties of these \LaTeX{} macros are of no concern to
the user, because they are designed to do their work quietly ``behind
the scenes''.

\subsection{\MF{} macros}

The \MF{} macros are much richer than their \LaTeX{} counterparts.
They have to deal with drawing primitives, linear algebra and abstract
representations of graphs.

\subsubsection{Transformers}

An important tool for generating complex graphs with arcs of different
styles is provided by transformers.  These are functions that take a
simple path (determined from the layout algorithm or specified
explicitly) as argument and return another path which corresponds to a
decorated version.  Here is an example that is used for
implementing gluon lines:
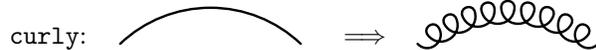
\begin{quote}
  |curly|:
    \begin{fmfgraph}(30,10)
      \fmfleft{i,d1} \fmfright{o,d2} \fmf{plain,left=0.4}{i,o}
    \end{fmfgraph} $\Longrightarrow$
    \begin{fmfgraph}(30,10)
      \fmfleft{i,d1} \fmfright{o,d2} \fmf{curly,left=0.4}{i,o}
    \end{fmfgraph}
\end{quote}
Using similar transformers, the implementation of dedicated drawing
functions is a matter of combining simple building blocks:
\begin{verbatim}
  style_def gluon_with_arrow expr p =
    draw (wiggly p);
    fill (arrow p)
  enddef;
\end{verbatim}
\fmfcmd{%
style_def charged_boson expr p =
  draw (wiggly p);
  fill (arrow p)
enddef;}
\begin{quote}
  \begin{fmfgraph}(30,10)
    \fmfleft{i,d1} \fmfright{o,d2} \fmf{plain,left=0.4}{i,o}
  \end{fmfgraph} $\Longrightarrow$
  \begin{fmfgraph}(30,10)
    \fmfleft{i,d1} \fmfright{o,d2} \fmf{wiggly,left=0.4}{i,o}
  \end{fmfgraph} $\Longrightarrow$
  \begin{fmfgraph}(30,10)
    \fmfleft{i,d1} \fmfright{o,d2} \fmf{charged_boson,left=0.4}{i,o}
  \end{fmfgraph}
\end{quote}
The current implementation does not attempt to force decorations
(e.g.~arrows) into the transformer paradigm, because \MF{} treats
drawing along a path differently from filling an outline.  Therefore
decorations are drawn after each other and not added to the object in
a pipeline.

\subsubsection{Graphs}

Graphs are represented by an array of vertices and arrays of vertices
emanating from the vertices.  Therefore the core of the data structure
for graphs is given by
\begin{verbatim}
  numeric vlist.first;
  numeric vlist.last;
  pair    vlist[]loc;
  numeric vlist[]arc.first;
  numeric vlist[]arc.last;
  numeric vlist[]arc[];
  numeric vlist[]arc[]tns;
\end{verbatim}
Here |vlist[|$i$|]loc|, with $\text{\texttt{vlist.first}} \le i \le
\text{\texttt{vlist.last}}$, is an array of two dimensional
coordinates, one for each vertex.  These coordinates start in the
state |unknown| and become |known| when the layout equations have been
solved.  For each vertex~$i$, |vlist[|$i$|]arc[|$j$|]|, with
$\text{\texttt{vlist[}$i$\texttt{]arc.first}} \le j \le
\text{\texttt{vlist[}$i$\texttt{]arc.last}}$, is an array of numbers
pointing to another vertex.  Therefore, each entry corresponds to an
arc.  The |vlist[|$i$|]arc[|$j$|]tns| array holds the
elements~$t_{ij}$ of the tension matrix.

This data structure is sufficient for performing the algorithmic
layout, as described below.  It is supplemented by similar arrays
holding information on linear constraints, line styles, etc.

\subsubsection{Linear algebra}

Let us now discuss how to solve the layout equation~(\ref{eq:layout})
for the common case of no constraints
\begin{equation}
\tag{\ref{eq:layout}$'$}
  0 = \sum_{\substack{i,j = 1\\ v_i \connected v_j}}^n
         t_{ij} (v_i^\mu - v_j^\mu)\,.
\end{equation}
Adding the constraints is a straightforward exercise, which is omitted
here for brevity.  \MF's
syntactical features and builtin linear algebra allow a direct
translation of~(\ref{eq:layout}$'$):
\begin{verbatim}
for i = vlist.first upto vlist.last:
  if unknown vlist[i]loc:
    origin = origin
    for j = vlist[i]arc.first upto vlist[i]arc.last:
      + vlist[i]arc[j]tns * (vlist[i]loc - vlist[vlist[i]arc[j]]loc)
    endfor;
  fi
endfor
\end{verbatim}
\MF's syntactical feature that allows this translation
of~(\ref{eq:layout}$'$) is the decoupling of control structures (|for|
\ldots |endfor|, |if| \ldots |fi|) from mathematical expressions.
This means that the bodies of loops and conditionals do not have to
form syntactically complete expressions.  We can therefore use loops
to construct expressions from building blocks.  For
the example of figure~\ref{fig:simple} the above fragment expands to
\begin{verbatim}
  origin = origin
    + vlist[5]arc[1]tns * (vlist[5]loc - vlist[1]loc)
    + vlist[5]arc[2]tns * (vlist[5]loc - vlist[2]loc)
    + vlist[5]arc[6]tns * (vlist[5]loc - vlist[6]loc);
  origin = origin
    + vlist[6]arc[3]tns * (vlist[6]loc - vlist[3]loc)
    + vlist[6]arc[4]tns * (vlist[6]loc - vlist[4]loc)
    + vlist[6]arc[5]tns * (vlist[6]loc - vlist[5]loc);
\end{verbatim}
since the vertices $v_1,v_2,v_3,v_4$ are already fixed as external
vertices.
If all tensions are unity, this is the linear system
\begin{equation}
\label{eq:simple}
  \begin{split}
    0 & = 3v_5 - v_1 - v_2 - v_6 \\
    0 & = 3v_6 - v_3 - v_4 - v_5
  \end{split}
\end{equation}
with unique solution
\begin{equation}
\tag{\ref{eq:simple}$'$}
  \begin{split}
    v_5 & = \frac{1}{8}\left(3v_1 + 3v_2 + v_3 + v_4\right) \\
    v_6 & = \frac{1}{8}\left(3v_3 + 3v_4 + v_1 + v_2\right)\,.
  \end{split}
\end{equation}
Note that we do not have to find the solution ourselves, because \MF{}
will not interpret the equations as assignments, but rather as linear
equations.  Once enough equations are given, the state of the vertex
coordinate will change from |unknown| to |known| and will have a
value.  As long as all vertices belong to a subgraph with at least one
element in~$V^{\text{ext}}$, there will be a unique solution to the
layout equations~(\ref{eq:layout}).

\subsubsection{Labels}

An interesting feature of \FMF{} is the ability to calculate optimal
label positions
in \MF{} and to communicate this information back to \LaTeX's
|picture| information.  Because \MF{} can not write any other files
than its |log|-file, the information has to be stored there and \TeX{}
macros have to be used to parse this file.

The algorithm used is quite simple.  It will place all labels on the
outside of the arc or vertex it is associated to.  If the result is
not satisfactory, explicit placement rules can be specified to
overwrite the automatic layout.

\subsubsection{Immediate mode}

The implementation of immediate mode is fairly straightforward,
because all necessary drawing primitives for drawing \FMF's line
styles on \MF{} paths are already available for graph mode.  Therefore
only trivial \LaTeX{} macros have to be written that translate the
\LaTeX{} syntax to \MF.

\subsubsection{Extension mechanism}

The extension mechanism serves two major purposes: it allows users to
specify new line styles and to overload existing line styles.

The ability to overload line styles can be used for a purely symbolic
markup of graphs.  If all the arcs are tagged by symbolic names like
|gluon|, |technipion|, |chargino|, etc., each user can use a library
of style definitions to render the graph in a customized visual
appearance.

\section{Usage}
\label{sec:usage}

Here is a short summary of the most important user commands of \FMF.
This section is not intended to replace the user's manual that comes
with the distribution~\cite{Ohl95a}, but it should give nevertheless
an idea how \FMF{} is used.

The overall structure is controlled by two environments:
\begin{commands}
  \item |\begin{fmffile}{|\meta{name}|}| \ldots |\end{fmffile}|\\
    This environment encloses all graphs that are written into a \MF{}
    input file named \meta{name}|.mf|.  For technical reasons,
    \meta{name} \emph{must not} be identical to the name of the main
    \LaTeX{} input file.  All created files have to be processed by
    \MF{} after the first run of \LaTeX{}.  See the \FMF{} user's
    manual for details on how to run \MF{} on various systems.
  \item |\begin{fmfgraph*}(|\meta{w}|,|\meta{h}|)|
      \ldots |\end{fmfgraph*}|\\
    This environment encloses a single graph of width \meta{w} and
    height \meta{h} \LaTeX{} |\unitlength|s.  There is also a light
    weight unstarred version that does not support labels.
\end{commands}
These environments can be used everywhere in a \LaTeX{} document, in
particular in centered |\parbox|es for
creating graphical equations like
\begin{verbatim}
  \parbox{20mm}{\begin{fmfgraph}(20,15)
    \fmfleft{i} \fmfright{o} \fmf{dashes}{i,v,v,o}
  \end{fmfgraph}} +
  \parbox{20mm}{\begin{fmfgraph}(20,15)
    \fmfleft{i} \fmfright{o} \fmf{dashes}{i,v1} \fmf{dashes}{v2,o}
    \fmf{fermion,left,tension=.3}{v1,v2,v1}
  \end{fmfgraph}} = \ln\Lambda^2
\end{verbatim}
\begin{displaymath}
  \parbox{20mm}{\begin{fmfgraph}(20,15)
    \fmfleft{i} \fmfright{o} \fmf{dashes}{i,v,v,o}
  \end{fmfgraph}} +
  \parbox{20mm}{\begin{fmfgraph}(20,15)
    \fmfleft{i} \fmfright{o} \fmf{dashes}{i,v1} \fmf{dashes}{v2,o}
    \fmf{fermion,left,tension=.3}{v1,v2,v1}
  \end{fmfgraph}} = \ln\Lambda^2
\end{displaymath}

\subsection{Graph mode}

The placement of external vertices is controlled by the following
commands:
\begin{commands}
  \item |\fmfleft{|\meta{$v_1$}$[$|,|\ldots$]$|}|\\
    place the vertices \meta{$v_1$},\ldots equidistantly on a smooth path
    on the left side of the diagram.
    There are analogous |\fmfright|, |\fmftop|, |\fmfbottom|, and
    |\fmfsurround| commands.  The latter will place the vertices on a
    smooth path surrounding the diagram.
  \item |\fmfleftn{|\meta{$v$}|}{|\meta{$n$}|}|\\
    place the vertices \meta{$v_1$},\ldots, \meta{$v_n$} equidistantly on
    a smooth path on the left side of the diagram.  Analogously for
    |\fmfrightn|, |\fmftopn|, |\fmfbottomn|, and |\fmfsurroundn|.
\end{commands}
The external vertices can be connected with themselves and internal
vertices by the commands
\begin{commands}
  \item |\fmf{|\meta{style}$[$|,|\meta{opt}$[$|,|\ldots$]]$|}{|%
      \meta{$v_1$}|,|\meta{$v_2$}$[$|,|\ldots$]$|}|\\
    connect the vertices $v_1$, $v_2$, \ldots by a line of style
    \meta{style} with options \meta{opt} switched on.  The
    available line styles are collected in
    table~\ref{tab:line-styles}.  Additional styles can be defined
    with |style_def|.  A special line style is |phantom|, which will
    not draw an arc at all.  It is nevertheless useful for
    manipulating the layout, because the corresponding arc enters the
    layout equations.  For convenience and for allowing more
    descriptive specifications, several aliases like |gluon|, |quark|,
    or |photon| of the line styles in table~\ref{tab:line-styles} are
    defined.
    Among the options, which can be given in a comma separated list
    after the line style are
    \begin{commands}
      \item |tension|: change the tension matrix element~$t_{ij}$ from
        the default value of 1.
      \item |left|, |right|: draw on a half circle on the left or right.
      \item |label|: arbitrary \TeX{} commands for labeling the arc
        (macros should be protected with |\noexpand|).
      \item |label.side|, |label.dist|: force placement of the label.
        on the |left| or |right| at this distance
      \item |width|: change the width of the line.
      \item |foreground|, |background|:
        colors (available with \MP{} only!).
    \end{commands}
  \item |\fmfn{|\meta{style}$[$|,|\meta{opt}$[$|,|\ldots$]]$|}{|%
      \meta{$v$}|}{|\meta{$n$}|}|\\
    connect the vertices $v_1$, \ldots, $v_n$ by a line of style
    \meta{style} with options \meta{opt} switched on.
\end{commands}
\DeleteShortVerb{\|}
\MakeShortVerb{\"}
\newcommand{\linesample}[1]{%
  \begin{fmfgraph}(30,4)
    \fmfleft{i}
    \fmfright{o}
    \fmf{#1}{i,o}
  \end{fmfgraph}}
\begin{table}[t]
  \hfuzz=3cm
  \begin{center}
  \begin{tabular}{|l|c|l|}\hline
    Name               &Example                &Parameters  \\\hline\hline
    "curly"            &\linesample{curly}           &"curly_len" \\\hline
    "dbl_curly"        &\linesample{dbl_curly}       &"curly_len" \\\hline
    "dashes"           &\linesample{dashes}          &"dash_len"  \\\hline
    "dashes_arrow"     &\linesample{dashes_arrow}    &"dash_len"  \\\hline
    "dbl_dashes"       &\linesample{dbl_dashes}      &"dash_len"  \\\hline
    "dbl_dashes_arrow" &\linesample{dbl_dashes_arrow}&"dash_len"  \\\hline
    "dots"             &\linesample{dots}            &"dot_len"   \\\hline
    "dots_arrow"       &\linesample{dots_arrow}      &"dot_len"   \\\hline
    "dbl_dots"         &\linesample{dbl_dots}        &"dot_len"   \\\hline
    "dbl_dots_arrow"   &\linesample{dbl_dots_arrow}  &"dot_len"   \\\hline
    "phantom"          &\linesample{phantom}         &            \\\hline
    "phantom_arrow"    &\linesample{phantom_arrow}   &            \\\hline
    "plain"            &\linesample{plain}           &            \\\hline
    "plain_arrow"      &\linesample{plain_arrow}     &            \\\hline
    "dbl_plain"        &\linesample{dbl_plain}       &            \\\hline
    "dbl_plain_arrow"  &\linesample{dbl_plain_arrow} &            \\\hline
    "wiggly"           &\linesample{wiggly}          &"wiggly_len"\\\hline
    "dbl_wiggly"       &\linesample{dbl_wiggly}      &"wiggly_len"\\\hline
    "zigzag"           &\linesample{zigzag}        &"zigzag_width"\\\hline
    "dbl_zigzag"       &\linesample{dbl_zigzag}      &"zigzag_len"\\\hline
  \end{tabular}
  \end{center}
  \caption{\label{tab:line-styles}Available line styles}
\end{table}
\DeleteShortVerb{\"}
\MakeShortVerb{\|}
The decoration of vertices is affected by the commands:
\begin{commands}
  \item |\fmfv{|\meta{opt}$[$|,|\ldots$]$|}{|%
      \meta{$v_1$}$[$|,|\ldots$]$|}|\\
    turn on the options \meta{opt} for the vertices $v_1$, \ldots.
    Among them are
    \begin{commands}
      \item |label|: arbitrary \TeX{} commands for labeling the vertex
        (macros should be protected with |\noexpand|).
      \item |label.angle|, |label.dist|: force placement of the label
        at this angle or distance.
      \item |decoration.size|, |decoration.filled|: size and filling
        style of the decoration.
      \item |decoration.shape|, |decoration.angle|: shape of the
        decoration, optionally rotated.
      \item |foreground|, |background|:
        colors (available with \MP{} only!).
    \end{commands}
    \def\VertexSample#1#2{%
      \parbox{8mm}{\begin{fmfgraph}(8,8)
        \fmfforce{(.5w,.4h)}{c}
        \fmfv{d.shape=#1,d.filled=#2,d.size=14pt}{c}
      \end{fmfgraph}}}
    Here are some examples for vertex decorations:
    \begin{commands}
      \item shaded (|fill=.5|) |circle| : \VertexSample{circle}{.5} and
        hatched (|-.5|) |square|: \VertexSample{square}{-.5}
      \item open (|fill=0|) |triangle|, |diamond|, |pentagon|, and
        |hexagon|:\\
        \VertexSample{triangle}{0} \qquad \VertexSample{diamond}{0} \qquad
        \VertexSample{pentagon}{0} \qquad \VertexSample{hexagon}{0}
      \item filled (|fill=1|) |triagram|, |tetragram|, |pentagram|,
        and |hexagram|:
        \VertexSample{triagram}{1} \qquad \VertexSample{tetragram}{1}
        \qquad \VertexSample{pentagram}{1}
        \qquad \VertexSample{hexagram}{1}
    \end{commands}
  \item |\fmfvn{|\meta{opt}$[$|,|\ldots$]$|}{|%
      \meta{$v$}|}{|\meta{$n$}|}|\\
    turn on the options \meta{opt} for the vertices $v_1$, \ldots,
    $v_n$.
  \item |\fmfblob{|\meta{d}|}{|\meta{$v_1$}$[$|,|\ldots$]$|}|\\
    draw a blob of diameter \meta{d} at the vertices $v_1$, \ldots.
    There is an analogous |\fmfblobn| command.
  \item |\fmfdot{|\meta{$v_1$}$[$|,|\ldots$]$|}|\\
    draw a dot at the vertices $v_1$, \ldots.
    There is an analogous |\fmfdotn| command.
  \item |\fmflabel{|\meta{label}|}{|\meta{$v_1$}$[$|,|\ldots$]$|}|\\
    label the vertices $v_1$, \ldots with \meta{label}.
\end{commands}
The location of vertices can be fixed, but experience shows that this
command should only be used as a last resort:
\begin{commands}
  \item |\fmfforce{|\meta{z}|}{|\meta{v1}\allowbreak$[$|,|\ldots$]$|}|\\
    place the vertices $v_1$, \ldots at the \MF{} coordinate \meta{z}.
\end{commands}
The layout calculation and drawing, which are implicitly performed at
the end of each |fmfgraph|, can be forced by
\begin{commands}
  \item |\fmffreeze|\\
    freeze the positions of the vertices entered so far.
  \item |\fmfdraw|\\
    draw all arcs and vertices entered so far.
\end{commands}
The appearance of the graph can be changed by
\begin{commands}
  \item |\fmfpen{|\meta{w}|}|\\
    change the width of the drawing ``pen'' to \meta{w}.  The default
    is $\mathop{\texttt{thin}}=1\mathop{\textrm{pt}}$.
  \item |\fmfset{|\meta{par}|}{|\meta{val}|}|\\
    set the parameter \meta{par} to the value \meta{val}.
  \item |\fmffixed{|\meta{d}|}{|\meta{v1}\allowbreak$[$|,|\ldots$]$|}|\\
    fix the distance vector between subsequent vertices in the list
    \meta{v1}, \ldots to \meta{d}.
\end{commands}

\subsection{Immediate mode}

In immediate mode, the drawing commands from graph mode are
duplicated with different arguments.  Therefore all decorated line
styles are available, but now for arbitrary \MF{} paths.
\begin{commands}
  \item |\fmfi{|\meta{style}$[$|,|\meta{opt}$[$|,|\ldots$]]$|}{|%
      \meta{path}|}|\\
    draw a line of style \meta{style} on the \MF{} path \meta{path}
    with options \meta{opt} switched on.
  \item |\fmfiv{|\meta{opt}$[$|,|\ldots$]$|}{|\meta{z}|}|\\
    draw a vertex with options \meta{opt} at the \MF{} coordinate $z$.
  \item |\fmfipair{|\meta{var}|}|, |\fmfipath{|\meta{var}|}|\\
    declare the variable \meta{var} as a pair (coordinate) or path.
  \item |\fmfiset{|\meta{x}|}{|\meta{y}|}|\\
    assign the value of \meta{y} to the variable \meta{x}.
  \item |\fmfiequ{|\meta{x}|}{|\meta{y}|}|\\
    declare equality of the variables \meta{x} and \meta{y}.  This is
    different from assignment, because \MF{} can solve linear
    equations~\cite{Knu86b}.
\end{commands}
Immediate and graph mode can be interfaced by using the following
\MF{} functions to access coordinates from graph mode in immediate mode:
\begin{commands}
  \item |vpath|$[$\meta{tag}$]$|(__|\meta{from}|,__|\meta{to}|)|\\
    return the \MF{} path of the arc from vertex \meta{from} to vertex
    \meta{to}.  The optional \meta{tag} can be used to disambiguate
    arcs connecting the same vertices.
  \item |vloc(__|\meta{v}|)|\\
    return the position of the vertex in \MF{} coordinates.
\end{commands}

\subsection{Extension mechanism}

A powerful, but dangerous command is
\begin{commands}
  \item |\fmfcmd{|\meta{\MF-expression}|}|\\
    write an arbitrary \meta{\MF-expression} to the \MF{} file.  No
    semicolon is appended.
\end{commands}
A recommended application of |\fmfcmd| is in defining new line styles
with
\begin{commands}
  \item |style_def |\meta{name}| expr p = | \ldots |enddef;|\\
    define a new line style and register it as \meta{name}.
\end{commands}
here is an example that will draw a cross at the center of the arc:
\begin{verbatim}
  \fmfcmd{%
    vardef cross_bar (expr p, len, ang) =
     ((-len/2,0)--(len/2,0))
        rotated (ang + angle direction length(p)/2 of p)
        shifted point length(p)/2 of p
    enddef;
    style_def crossed expr p =
      cdraw p;
      ccutdraw cross_bar (p, 5mm,  45);
      ccutdraw cross_bar (p, 5mm, -45)
    enddef;}
\end{verbatim}
\fmfcmd{%
  vardef cross_bar (expr p, len, ang) =
   ((-len/2,0)--(len/2,0))
      rotated (ang + angle direction length(p)/2 of p)
      shifted point length(p)/2 of p
  enddef;
  style_def crossed expr p =
    cdraw p;
    ccutdraw cross_bar (p, 5mm,  45);
    ccutdraw cross_bar (p, 5mm, -45)
  enddef;}
as in
\begin{quote}
  |\fmfcmd{crossed}{v1,v2}|
    $\Longrightarrow$ \parbox{30mm}{\linesample{crossed}}
\end{quote}

\subsection{Advanced example}

\begin{figure}
  \begin{center}
  \begin{fmfgraph*}(50,30)
    \fmfpen{thick} \fmfleft{i} \fmfright{o}
    \fmf{boson,tension=2}{i,iv3} \fmf{boson,tension=2}{o,ov3}
    \fmf{quark}{iv1,iv2,iv3,iv1} \fmf{quark}{ov1,ov2,ov3,ov1}
    \fmf{gluon,tension=.5}{ov1,iv1} \fmf{gluon,tension=.5}{iv2,ov2}
    \fmfv{decor.shape=square,decor.size=4thick}{iv3,ov3}
    \fmfdot{iv1,iv2,ov1,ov2}
    \fmffixed{(0,.7h)}{iv1,iv2} \fmffixed{(0,.7h)}{ov1,ov2}
    \fmffreeze
    \fmfcmd{%
      save loc, bmin, bmax;
      forsuffixes $ = 1, 2, 3:
        (loc.$.x, loc.$.y) = vloc __iv.$;
      endfor
      bmax.x = max (loc1x, loc2x, loc3x) + .1w;
      bmax.y = max (loc1y, loc2y, loc3y) + .1h;
      bmin.x = min (loc1x, loc2x, loc3x) - .1w;
      bmin.y = min (loc1y, loc2y, loc3y) - .1h;}
    \fmfi{dashes,width=thin}{%
      (bmin.x,bmin.y) -- (bmax.x,bmin.y)
       -- (bmax.x,bmax.y) -- (bmin.x,bmax.y) -- cycle}
    \fmfiv{label=$\Gamma^5_{\mu\alpha\beta}$}%
      {(.5[bmin.x,bmax.x],bmax.y)}
  \end{fmfgraph*}
  \end{center}
  \caption{\label{fig:3loop}%
    Three loop QCD correction to the axial vector part of the
    $Z^0$-selfenergy.}
\end{figure}
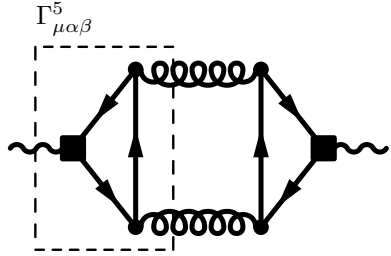
Finally, figure~\ref{fig:3loop} is a slightly more advanced example: a
three loop QCD correction to the axial vector part of the
$Z^0$-selfenergy. The beginning is straightforward: we connect the
vertices.  It is beneficial to give a higher tension to the |boson|s
and a lower tension to the |gluon|s to balance the diagram:
\begin{verbatim}
  \fmfpen{thick} \fmfleft{i} \fmfright{o}
  \fmf{boson,tension=2}{i,iv3} \fmf{boson,tension=2}{o,ov3}
  \fmf{quark}{iv1,iv2,iv3,iv1} \fmf{quark}{ov1,ov2,ov3,ov1}
  \fmf{gluon,tension=.5}{ov1,iv1} \fmf{gluon,tension=.5}{iv2,ov2}
\end{verbatim}
Now we add dots to the vector vertices and big squares to the axial
vector vertices:
\begin{verbatim}
  \fmfv{decor.shape=square,decor.size=4thick}{iv3,ov3}
  \fmfdot{iv1,iv2,ov1,ov2}
\end{verbatim}
As it stands, the diagram will have all vertices on a straight line.
To remedy this situations, we use |\fmffixed| to open the triangles:
\begin{verbatim}
  \fmffixed{(0,.7h)}{iv1,iv2} \fmffixed{(0,.7h)}{ov1,ov2}
\end{verbatim}
We could also have used |phantom| arcs to achieve similar results, but
here the linear constraints are more concise and intuitive.

For illustration, we mark the left triangle subgraph by a dashed box.
As of version \Version, \FMF{} does not have a builtin function for
calculating the enclosing box of a list of vertices.  However, this is
easily done in a few lines of \MF.  Before we start, we have to force
the layout calculation:
\begin{verbatim}
  \fmffreeze
\end{verbatim}
For convenience, we store the coordinates of the vertices in temporary
variables:
\begin{verbatim}
  \fmfcmd{%
    save loc, bmin, bmax;
    forsuffixes $ = 1, 2, 3:
      (loc.$.x, loc.$.y) = vloc __iv.$;
    endfor
\end{verbatim}
It is trivial to get the coordinates of the enclosing box:
\begin{verbatim}
    bmax.x = max (loc1x, loc2x, loc3x) + .1w;
    bmax.y = max (loc1y, loc2y, loc3y) + .1h;
    bmin.x = min (loc1x, loc2x, loc3x) - .1w;
    bmin.y = min (loc1y, loc2y, loc3y) - .1h;}
\end{verbatim}
Now we can use immediate mode to draw this box, thinly dashed
and labeled
\begin{verbatim}
  \fmfi{dashes,width=thin}{%
    (bmin.x,bmin.y) -- (bmax.x,bmin.y)
     -- (bmax.x,bmax.y) -- (bmin.x,bmax.y) -- cycle}
  \fmfiv{label=$\Gamma^5_{\mu\alpha\beta}$}%
    {(.5[bmin.x,bmax.x],bmax.y)}
\end{verbatim}

\section{Conclusions}
\label{sec:concl}

I have described \FMF, a flexible tool for portable and convenient
inclusion of Feynman diagrams in \LaTeX{} documents.

\section*{Acknowledgments}
\label{sec:ack}

I am most grateful to Wolfgang Kilian, who pushed \FMF's predecessor
|feynman.mf| to its limits in his doctoral thesis and provided
invaluable input for \FMF.  Thanks also to all brave users who tested
preview versions and provided encouragement.

\appendix
\section{Distribution}

The latest release of \FMF{} is available by anonymous ftp from
\begin{quote}
  |crunch.ikp.physik.th-darmstadt.de|
\end{quote}
in the directory
\begin{quote}
  |pub/ohl/feynmf|
\end{quote}
or from any of the Comprehensive \TeX{} Archive Network (CTAN) hosts
\begin{quote}
  |ftp.shsu.edu|, |ftp.tex.ac.uk|, |ftp.dante.de|
\end{quote}
in the directory
\begin{quote}
  |macros/latex/contrib/supported/feynmf|
\end{quote}
Important announcements (new versions, fatal bugs, etc.)
will be made through the mailing list
\begin{quote}
  |feynmf-announce@crunch.ikp.physik.th-darmstadt.de|
\end{quote}
Subscriptions can be obtained from
\begin{quote}
  |majordomo@crunch.ikp.physik.th-darmstadt.de|
\end{quote}
(send a message consisting of |help| to |majordomo| for
instructions on how to subscribe, don't send such messages to the list
itself).

\section{Installation}

\FMF{} comes in standard \LaTeX{} |doc| format~\cite{GMS94}.  The
installation procedure is described in the |README| file and need not
be repeated here.

\section{Revision History}
\label{sec:history}

\subsection*{Version 1.0, May 1995}
First official release.

\bibliography{journals-abbrev,fmfcpc}
\end{fmffile}
\end{document}